\documentclass{article}%
\usepackage{graphicx}
\usepackage{amsmath}
\usepackage{amsfonts}
\usepackage{amssymb}%
\providecommand{\U}[1]{\protect\rule{.1in}{.1in}}

\begin{document}

\author{Antony Valentini\\Augustus College}

\begin{center}
{\LARGE Long-time relaxation in pilot-wave theory}

\bigskip

\bigskip

\bigskip

\bigskip\ 

Eitan Abraham

\textit{Institute of Biological Chemistry, Biophysics and Bioengineering,}

\textit{School of Engineering and Physical Sciences, Heriot-Watt University,}

\textit{Edinburgh EH14 4AS, United Kingdom.}

\bigskip\ 

Samuel Colin, Antony Valentini

\textit{Department of Physics and Astronomy,}

\textit{Clemson University, Kinard Laboratory,}

\textit{Clemson, SC 29634-0978, USA.}

\bigskip

\bigskip

\bigskip

\bigskip
\end{center}

We initiate the study of relaxation to quantum equilibrium over long
timescales in pilot-wave theory. We simulate the time evolution of the
coarse-grained $H$-function $\bar{H}(t)$ for a two-dimensional harmonic
oscillator. For a (periodic) wave function that is a superposition of the
first 25 energy states we confirm an approximately exponential decay of
$\bar{H}$ over five periods. For a superposition of only the first four energy
states we are able to calculate $\bar{H}(t)$ over 50 periods. We find that,
depending on the set of phases in the initial wave function, $\bar{H}$ can
decay to a large nonequilibrium residue exceeding $10\%$ of its initial value
or it can become indistinguishable from zero (the equilibrium value). We show
that a large residue in $\bar{H}$ is caused by a tendency for the trajectories
to be confined to sub-regions of configuration space for some wave functions,
and that this is less likely to occur for larger numbers of energy states (if
the initial phases are chosen randomly). Possible cosmological implications
are briefly discussed.

\bigskip

\bigskip

\bigskip

\bigskip

\bigskip

\bigskip

\bigskip

\bigskip

\bigskip

\bigskip

\bigskip

\bigskip

\bigskip

\bigskip

\bigskip

\bigskip

\bigskip

\bigskip

\bigskip

\bigskip

\bigskip

\bigskip

\bigskip

\bigskip

\bigskip

\section{Introduction}

In the de Broglie-Bohm pilot-wave formulation of quantum theory
\cite{deB28,BV09,B52a,B52b,Holl93}, a system has an actual configuration
$q(t)$ which evolves in time with a velocity $\dot{q}\equiv dq/dt$ that is
determined by the wave function $\psi(q,t)$, where $\psi$ satisfies the
Schr\"{o}dinger equation $i\partial\psi/\partial t=\hat{H}\psi$ (taking
$\hbar=1$). For standard Hamiltonians, $\dot{q}$ is proportional to the
gradient $\partial_{q}S$ of the phase $S$ of $\psi$. Generally, $\dot
{q}=j/|\psi|^{2}$ where $j=j\left[  \psi\right]  =j(q,t)$ is the
Schr\"{o}dinger current \cite{SV08}. Here $\psi$ is a `pilot wave' in
configuration space that guides the motion of an individual system; it has no
\textit{a priori} connection with probabilities.

For an ensemble of systems with the same initial wave function $\psi(q,t_{i}%
)$, we may consider an arbitrary initial distribution $\rho(q,t_{i})$ of
configurations $q(t_{i})$. The time evolution $\rho(q,t)$ of the distribution
will be determined by the continuity equation%
\begin{equation}
\frac{\partial\rho}{\partial t}+\partial_{q}\cdot\left(  \rho\dot{q}\right)
=0\ .
\end{equation}
Because $\left\vert \psi\right\vert ^{2}$ obeys the same continuity equation,
an initial distribution $\rho(q,t_{i})=\left\vert \psi(q,t_{i})\right\vert
^{2}$ will evolve into $\rho(q,t)=\left\vert \psi(q,t)\right\vert ^{2}$. In
this state of `quantum equilibrium' the probability distribution matches the
Born rule. But in pilot-wave theory one may just as well consider
`nonequilibrium' distributions $\rho(q,t_{i})\neq\left\vert \psi
(q,t_{i})\right\vert ^{2}$ (just as in classical mechanics one may consider
non-thermal distributions) \cite{AV91a,AV91b,AV92}.

As is well known, pilot-wave dynamics reproduces the empirical predictions of
quantum theory if the initial ensemble is in quantum equilibrium
($\rho(q,t_{i})=\left\vert \psi(q,t_{i})\right\vert ^{2}$) \cite{B52a,B52b}.
However, for an initial nonequilibrium ensemble ($\rho(q,t_{i})\neq\left\vert
\psi(q,t_{i})\right\vert ^{2}$), the statistical predictions will generally
disagree with quantum theory. Quantum physics may then be regarded as a
special equilibrium case of a wider nonequilibrium physics
\cite{AV91a,AV91b,AV92,AV96,AV01,AV02,AV07,AV08,AV09,AV10,PV06}.

The quantum-theoretical equilibrium state $\rho_{\mathrm{QT}}=\left\vert
\psi\right\vert ^{2}$ may be seen to arise from a process of relaxation
analogous to classical thermal relaxation. The approach to equilibrium may be
quantified by the coarse-grained $H$-function%
\begin{equation}
\bar{H}=\int dq\ \bar{\rho}\ln(\bar{\rho}/\bar{\rho}_{\mathrm{QT}})\ ,
\label{Hbar}%
\end{equation}
where $\bar{\rho}$, $\bar{\rho}_{\mathrm{QT}}$ are respectively obtained by
averaging $\rho$, $\rho_{\mathrm{QT}}$ over (non-overlapping) coarse-graining
cells. This function obeys a coarse-graining $H$-theorem $\bar{H}(t)\leq
\bar{H}(0)$ (assuming that the initial state at $t=0$ has no fine-grained
micro-structure), where the minimum $\bar{H}=0$ corresponds to equilibrium
$\bar{\rho}=\bar{\rho}_{\mathrm{QT}}$ \cite{AV91a,AV92,AV01}. Like the
analogous classical result, this theorem provides a general mechanism in terms
of which one can understand how equilibrium is approached, while not proving
that equilibrium is actually reached. The extent to which relaxation occurs
will depend on the system and on the initial conditions. Extensive numerical
simulations have been performed for initial wave functions that are
superpositions of energy eigenstates. It has been found that initial
nonequilibrium distributions $\rho$ rapidly approach $\rho_{\mathrm{QT}}$ on a
coarse-grained level (that is, $\bar{\rho}\longrightarrow\bar{\rho
}_{\mathrm{QT}}$) \cite{AV92,AV01,VW05,TRV12,SC12}, with an approximately
exponential decay of $\bar{H}(t)$ with time \cite{VW05,TRV12}.

The rapid relaxation seen in these simulations suggests that we may understand
the Born rule as having arisen from a relaxation process that presumably took
place in the remote past. Quantum nonequilibrium may have existed in the early
universe, at very early times before relaxation took place
\cite{AV91a,AV91b,AV92,AV96}. This could have left observable traces today --
in the cosmic microwave background (CMB) or in relic particles that decoupled
at very early times \cite{AV01,AV07,AV08,AV09,AV10,CV13}. But for ordinary
laboratory systems one may reasonably expect to find quantum equilibrium today
to high accuracy (as has been experimentally confirmed in a wide range of
conditions). This expectation rests on the assumption that over long
timescales relaxation will continue until any residual nonequilibrium becomes
very small \cite{AV92}. Given the results of numerical simulations carried out
so far, and given the extremely violent history of our universe, this
assumption seems reasonable. Still, the assumption should be tested with more
extensive simulations that aim at probing relaxation in the long-time limit.
One of the purposes of this paper is to initiate such a study.

Another point that requires further study is the generality of the relaxation
process. As in classical statistical mechanics, there will always be initial
states that do not relax to equilibrium. For example, the ground state of a
box or oscillator has vanishing de Broglie velocity ($\partial_{q}S=0$) so
that any initial nonequilibrium distribution will be static and cannot relax.
But to understand the quantum probabilities that we observe today, we must
bear in mind that every system we have access to has a long and violent
astrophysical history -- during which its quantum state will have been a
complicated superposition of many energy states. What matters is what
realistically might have occurred in the remote past in our actual universe.
Could there exist realistic early states that do not relax?

Contopoulos \textit{et al}. \cite{CDE12} have shown that, for a
two-dimensional oscillator in a superposition of three or four energy states,
there are some choices of initial wave function for which relaxation is
limited and may hardly occur at all (see also ref. \cite{EC06}). In these
cases the trajectories are confined to sub-regions of configuration space and
do not explore the whole support of $\left\vert \psi\right\vert ^{2}$. It was
found that this can occur regardless of whether or not the trajectories are
chaotic. In some of the examples studied the trajectories are excluded from a
small sub-region so that relaxation does not occur completely; in other
examples the trajectories are confined to a small sub-region so that there is
hardly any relaxation. The behaviour of the trajectories depends on the
initial wave function (which determines the velocity field); it also depends
on the support of the initial distribution (which determines where the
trajectories begin). A similar example was given by Colin \cite{SC12} for a
Dirac fermion in a spherical box: for certain superpositions of three energy
states trajectories are trapped inside a `core' close to the origin, so that
initial distributions confined to the core do not relax. These examples raise
the question of how common such confinement is, and whether states of the
early universe are likely to exhibit such behaviour.

Inflationary cosmology opens an empirical window onto quantum probabilities in
the very early universe. According to our best current understanding, the
observed temperature anisotropy of the CMB was seeded by quantum fluctuations
in a scalar field during an early period of inflationary expansion
\cite{LL00,Muk05,W08,PU09}. It has been shown that quantum nonequilibrium
during the inflationary phase could leave an observable imprint today in the
CMB \cite{AV07,AV08,AV10,CV13}. Thus the possibility of quantum nonequilibrium
in the early universe can be tested using inflationary cosmology.

It has also been shown that relaxation can be suppressed at long wavelengths
on expanding space \cite{AV07,AV08,AV10,CV13,AVbook}. In a cosmology with a
`pre-inflationary' phase -- for example a period of radiation-dominated
expansion, prior to inflation \cite{VF82,L82,S82,PK07,WN08} -- one may then
expect to find a large-scale power deficit in the CMB (above some critical
wavelength) \cite{AV07,AV08,AV10,CV13}. Such a deficit has recently been found
in data from the \textit{Planck} satellite \cite{PlanckXV}. It is conceivable
that the observed deficit is caused by relaxation suppression during a
pre-inflationary era, though of course it might also be due to some other
effect. (For a detailed discussion see refs. \cite{AV10,CV13}.)

It may also be of interest to consider a cosmological pre-inflationary phase
with a very small number of excitations above the vacuum, since during
inflation itself it is generally assumed that the quantum state is in or very
close to the vacuum. If the pre-inflationary phase contained a very small
number of excitations, one would like to know whether or not equilibrium is to
be expected at the onset of inflation itself (and hence whether to expect
anomalies in the CMB). This motivates us to study relaxation for states
containing a small number of energy states -- another focus of the present paper.

It suffices to consider the two-dimensional harmonic oscillator, since this
system arises in high-energy field theory and cosmology when one considers a
decoupled (that is, unentangled) field mode. To see this, consider a free,
minimally-coupled, and massless scalar field $\phi$ on an expanding flat space
with scale factor $a(t)$ (where $t$ is standard cosmological time). Working in
Fourier space, with field components $\phi_{\mathbf{k}}(t)$, physical
wavelengths are proportional to $a(t)$. If we write $\phi_{\mathbf{k}}$ in
terms of its real and imaginary parts, $\phi_{\mathbf{k}}=\frac{\sqrt{V}%
}{(2\pi)^{3/2}}\left(  q_{\mathbf{k}1}+iq_{\mathbf{k}2}\right)  $ (where $V$
is a normalisation volume), in terms of the real variables $q_{\mathbf{k}r}$
the field Hamiltonian becomes a sum $H=\sum_{\mathbf{k}r}H_{\mathbf{k}r}$,
where $H_{\mathbf{k}r}$ is the Hamiltonian of a harmonic oscillator with mass
$m=a^{3}$ and angular frequency $\omega=k/a$ \cite{AV07,AV08,AV10}. An
unentangled mode $\mathbf{k}$ has an independent dynamics, with wave function
$\psi_{\mathbf{k}}(q_{\mathbf{k}1},q_{\mathbf{k}2},t)$. Dropping the index
$\mathbf{k}$, the wave function $\psi=\psi(q_{1},q_{2},t)$ satisfies the
Schr\"{o}dinger equation%
\begin{equation}
i\frac{\partial\psi}{\partial t}=\sum_{r=1,\ 2}\left(  -\frac{1}{2m}%
\partial_{r}^{2}+\frac{1}{2}m\omega^{2}q_{r}^{2}\right)  \psi\ , \label{S2D'}%
\end{equation}
while de Broglie's equation of motion for the configuration $(q_{1},q_{2})$
reads%
\begin{equation}
\dot{q}_{r}=\frac{1}{m}\operatorname{Im}\frac{\partial_{r}\psi}{\psi}
\label{deB2D'}%
\end{equation}
(with $\partial_{r}\equiv\partial/\partial q_{r}$). The marginal distribution
$\rho=\rho(q_{1},q_{2},t)$ for the mode evolves according to%
\begin{equation}
\frac{\partial\rho}{\partial t}+\sum_{r=1,\ 2}\partial_{r}\left(  \rho\frac
{1}{m}\operatorname{Im}\frac{\partial_{r}\psi}{\psi}\right)  =0\ .
\label{C2D'}%
\end{equation}
Equations (\ref{S2D'})--(\ref{C2D'}) are identical to those of pilot-wave
dynamics for a nonrelativistic two-dimensional harmonic oscillator with a
time-dependent mass $m=a^{3}$ and time-dependent angular frequency
$\omega=k/a$. Thus we may discuss relaxation for a single field mode in terms
of relaxation for a nonrelativistic harmonic oscillator \cite{AV07,AV08}.

It has been shown that the time evolution defined by equations (\ref{S2D'}%
)--(\ref{C2D'}) is mathematically equivalent to the time evolution of a
standard harmonic oscillator (with constant mass $m$ and constant angular
frequency $\omega$) but with real time replaced by a `retarded time' that
depends on the wavelength of the mode \cite{CV13}. This result shows that
cosmological relaxation for a single field mode may be discussed in terms of
relaxation for a standard harmonic oscillator. (In the short-wavelength limit
the retarded time reduces to real time and we recover the time evolution of a
field mode on Minkowski spacetime. In this limit, if the field mode is in a
superposition of different states of definite occupation number, then
relaxation will take place as for an ordinary oscillator. On the other hand,
at long wavelengths -- roughly speaking, for physical wavelengths larger than
the Hubble radius $H^{-1}\equiv a/\dot{a}$ -- relaxation is retarded. See ref.
\cite{CV13}.)

In this paper we therefore study relaxation for the standard two-dimensional
harmonic oscillator -- in the long-time limit and also with a small number of
excitations. We calculate as far ahead in time as we are able to guarantee
accurate results. We pay particular attention to the accuracy of our
calculations of $\bar{H}(t)$, with a view to quantifying the precise extent of
relaxation in the long-time limit. We are especially interested in the
following questions. For a given initial wave function, does $\bar{H}(t)$
continue to decrease approximately exponentially for an indefinite period of
time or does the decay eventually halt? If there are only a tiny number of
energy states in the superposition, can the resulting very slow evolution
nevertheless eventually drive the system to equilibrium to high accuracy? To
what extent does the behaviour of $\bar{H}(t)$ depend on the details of the
initial quantum state?

We first consider (Section 3) a case where the wave function is a
superposition of 25 energy states. The approximate exponential decay of
$\bar{H}(t)$ is found to continue for as far in time as we are able to
simulate reliably, with $\bar{H}(t)$ reaching a final value that is smaller
than $1\%$ of its initial value. We then consider (Section 4) a case where the
wave function is a superposition of four energy states. We find that $\bar
{H}(t)$ decays approximately exponentially at first but eventually the decay
halts and the function $\bar{H}(t)$ levels off to a roughly constant `residue'
that exceeds $10\%$ of the initial value. To explain the difference between
these two cases, we consider (Section 5) the extent to which the trajectories
explore the full support of $\left\vert \psi\right\vert ^{2}$. For the first
case we find negligible confinement to sub-regions, while for the second case
we find strong confinement. Thus we may explain the large residue in $\bar
{H}(t)$ for the second case as due to a lack of full exploration of the
support of $\left\vert \psi\right\vert ^{2}$ by the trajectories, along the
lines found by Contopoulos \textit{et al}. \cite{CDE12}. We also consider how
common these different kinds of behaviour are likely to be. While a full
answer is left to future research, we present evidence that confinement of
trajectories -- and an associated large residue in $\bar{H}$ -- are less
likely to occur when the initial wave function contains a larger number of
superposed energy states. This suggests that a large nonequilibrium residue,
as quantified by a large non-zero value of $\bar{H}$, is likely to be
cosmologically relevant only in scenarios containing a very small number of
energy states (for example, during pre-inflation).

Before proceeding let us summarise previous numerical work in which the time
evolution of $\bar{H}$ was studied.\footnote{It should be noted that while
simulations of the time-evolving distribution $\rho(q,t)$ are fairly
straightforward to carry out, to obtain accurate values of $\bar{H}(t)$ is
computationally very demanding owing to the extremely irregular fine-grained
structure that develops in $\rho$ as a function of position $q$ in
configuration space.} For all the systems studied so far, including that in
this paper, the wave function is periodic in time. In refs. \cite{AV92,AV01}
numerical simulations were carried out for the artificial case of a
one-dimensional box, for which relaxation can take place only to a very
limited degree. For a wave function that is a superposition of the first 10
energy states, values of $\bar{H}(t)$ were obtained over just $1\%$ of a
period, with an initial value $\simeq0.61$ and a final value $\simeq0.56$, so
that $\bar{H}$ dropped only slightly -- to about $90\%$ of its initial value.
In ref. \cite{VW05} simulations were carried out for a two-dimensional box
with a superposition of the first 16 energy states. Values of $\bar{H}(t)$
were obtained over half a period, with an initial value $\simeq0.74$ and a
final value $\simeq0.18$, so that $\bar{H}$ dropped to about $25\%$ of its
initial value. In ref. \cite{TRV12} further simulations were carried out for
the same system, for varying numbers of energy states superposed in the wave
function. Values of $\bar{H}(t)$ were obtained over a whole period. For
example, for $16$ energy states it was found that an initial value
$\simeq0.84$ decreased to $\simeq0.08$ -- that is, $\bar{H}$ dropped to about
$10\%$ of its initial value.\footnote{In ref. \cite{CS10}, simulations were
carried out for the two-dimensional box with the wave function a superposition
of just 4 energy states. Relaxation was studied over two periods. In one
example $\bar{H}$ dropped to about $40\%$ of its initial value; however the
reported (very large) magnitudes of $\bar{H}$ were incorrectly normalised.}
Finally, in ref. \cite{CV13} simulations were carried out for the standard
oscillator with a superposition of the first 25 energy states. (See the case
of no spatial expansion, shown in figures 3 and 4 of ref. \cite{CV13}.) Values
of $\bar{H}(t)$ were obtained over five periods, with an initial value
$\simeq1.27$ and a final value $\simeq0.07$, so that $\bar{H}$ dropped to
about $5\%$ of its initial value. In refs. \cite{VW05,TRV12} an approximate
exponential decay $\bar{H}(t)\approx\bar{H}(0)\exp(-t/\tau)$ was observed over
the whole of the time intervals considered (where $\tau$ is the relaxation
timescale and $1/\tau$ is the decay rate). In ref. \cite{CV13}, over the five
periods considered there was a clear exponential decay during the first two or
three periods after which the decay rate $1/\tau$ appeared to diminish slightly.

\section{Set up and method}

We consider the two-dimensional harmonic oscillator (with constant mass and
angular frequency). We employ units such that $\hbar=m=\omega=1$.

The wave function is taken to be a superposition%
\begin{equation}
\psi(q_{1},q_{2},t)=\frac{1}{\sqrt{M}}\sum_{m,n=0}^{\sqrt{M}-1}e^{i\theta
_{mn}}e^{-i(m+n+1)t}\phi_{m}(q_{1})\phi_{n}(q_{2})\label{supn}%
\end{equation}
of the first $M$ energy eigenstates,\footnote{Note that $M$ is the number of
energy eigenfunctions or modes in the superposition, not the number of total
energy eigenvalues.} with normalised eigenfunctions%
\begin{equation}
\phi_{m}(q_{r})=\frac{1}{\pi^{1/4}}\frac{1}{\sqrt{2^{m}m!}}\mathcal{H}%
_{m}(q_{r})e^{-q_{r}^{2}/2}%
\end{equation}
(where $\mathcal{H}_{m}$ is the Hermite polynomial of order $m$). As in
previous studies the superposition is equally weighted, with randomly-chosen
initial phases $\theta_{mn}$, and the wave function is periodic in time (with
period $2\pi$). We shall first consider the case $M=25$ and then the case
$M=4$.\footnote{For each given $M$, the initial phases are chosen randomly but
then kept the same for all runs with the same $M$.}

In all the simulations reported in this paper we take the initial probability
density to be equal to the equilibrium density of the ground state:%
\[
\rho(q_{1},q_{2},0)=\left\vert \phi_{0}(q_{1})\phi_{0}(q_{2})\right\vert
^{2}\ .
\]
This choice is made purely on grounds of simplicity. A similar choice was made
for previous simulations in two dimensions. (The exploration of alternative
choices is left for future work.) Clearly $\rho(q_{1},q_{2},0)\neq
\rho_{\mathrm{QT}}(q_{1},q_{2},0)$ and the initial state is far from equilibrium.

To obtain the time evolution of the density $\rho(q_{1},q_{2},t)$, instead of
integrating the continuity equation (\ref{C2D'}) directly we employ the
`backtracking' method developed in ref. \cite{VW05}. The conserved ratio
$\rho/\rho_{\mathrm{QT}}$ along trajectories is used to construct $\rho$ on a
uniform grid at each time $t$, where each point on the grid is backtracked
along a trajectory to $t=0$ to find the required value of $\rho/\rho
_{\mathrm{QT}}$ (where the function $\rho/\rho_{\mathrm{QT}}$ is known
analytically at $t=0$ and $\rho_{\mathrm{QT}}$ is known analytically at all
times). We impose a precision of $0.025$ on each backtracked trajectory. This
is to be compared with the lengthscale $\sim1$ down to which the equilibrium
density $\rho_{\mathrm{QT}}$ displays structure. In sharp contrast, the
evolving nonequilibrium density $\rho$ develops an extremely irregular
fine-grained structure, varying rapidly on tiny lengthscales. (For a close-up
example, see figure 6 of ref. \cite{VW05}.)

The densities $\rho$, $\rho_{\mathrm{QT}}$ may be averaged over
non-overlapping coarse-graining cells, yielding coarse-grained values
$\bar{\rho}$, $\bar{\rho}_{\mathrm{QT}}$ which may be assigned to the centres
of each cell. These values are used to calculate the coarse-grained
$H$-function%
\begin{equation}
\bar{H}=\int\int dq_{1}dq_{2}\ \bar{\rho}\ln(\bar{\rho}/\bar{\rho
}_{\mathrm{QT}})\ . \label{Hbarint}%
\end{equation}

To display the densities themselves, it is convenient to plot `smoothed'
densities $\tilde{\rho}$, $\tilde{\rho}_{\mathrm{QT}}$ obtained by
coarse-graining with overlapping cells \cite{VW05}.\footnote{Our density plots
employ $76\times76$ overlapping cells each with $30\times30$ grid points. The
cells have side $\varepsilon=5/8$. For a given cell, shifting it along either
axis by a distance equal to $20\%$ of $\varepsilon$ generates a neighbouring
cell.}

To calculate the backtracked trajectories we employ the Runge-Kutta-Fehlberg
method with adaptive time-step (as used in refs. \cite{VW05,SC12}). The de
Broglie velocity field varies extremely rapidly in certain regions of
configuration space, in particular near nodes of the wave function. For this
reason, some trajectories are much harder to calculate than others. In fact,
there are usually a few trajectories for which the calculation would take an
unduly large number of computational steps. We set the maximum number of steps
per trajectory at $10^{7}$. (We divide the time interval into ten equal and
successive sub-intervals, and for each sub-interval the maximum number of
iterations is set to $10^{6}$.) Should this number of steps be exceeded the
trajectory is discarded, resulting in a point of the grid to which no value of
$\rho$ has been assigned. As long as the number of such points is small, the
overall simulation will still be accurate. We abort the simulation when the
percentage of accurate trajectories (successfully backtracked to $t=0$ from
the uniform grid at time $t$) drops below 95\%. With these restrictions we are
able to perform the simulations up to a maximum of five time periods (that is,
up to $t=10\pi$) for the case $M=25$ and a maximum of 50 time periods (that
is, up to $t=100\pi$) for the case $M=4$. With our current methods, when we
attempt to simulate these systems for longer times the percentage of accurate
trajectories drops below what we deem to be the minimal acceptable level of 95\%.

To evaluate the integral (\ref{Hbarint}) for $\bar{H}(t)$ we take $16\times16$
non-overlapping coarse-graining cells each of side $5/8$. The integration is
thus restricted to a box of side $10$ (centred on the origin), which exceeds
the linear scale of the discernible supports of the densities by a factor
$\sim2$.

It is surprisingly difficult to evaluate $\bar{H}$ accurately. Because of the
extremely irregular small-scale structure of the exact nonequilibrium density
$\rho$, the coarse-grained value $\bar{\rho}$ assigned to a cell can depend
significantly on the grid of points used to sample the (rapidly-varying)
function $\rho$ within the cell \cite{VW05}. To estimate the accuracy of
$\bar{H}$ we perform three separate simulations with three different grids:
the first grid has $29\times29$ points per cell, the second has $30\times30$
points per cell, and the third has $31\times31$ points per cell.\footnote{Note
that the second case, for example, employs a total grid of $480\times
480=230,400$ points.} Thus we sample the fine-grained function $\rho$ in three
different ways, yielding three different values for $\bar{H}$ at each time $t$.

As a further check we have also calculated the time evolution of $\rho$ by
evolving an initially uniform grid of points forwards in time, using an
independently written code. If the number of initial points is large enough,
then despite the clustering around the maxima of $\rho_{\mathrm{QT}}$ --
resulting in a highly non-uniform grid at later times, with relatively sparse
regions around minima of $\rho_{\mathrm{QT}}$ -- one may still (by
straightforward interpolation) generate a smoothed distribution at later times
which by eye is hardly distinguishable from the results displayed here. This
provides a useful cross check on the validity of our simulations. However, to
guarantee accurate values for $\bar{H}$ we exclusively employ the (rather more
involved) backtracking method.

\section{An example of long-time relaxation}

We first consider a case with 25 energy states, $M=25$. As noted above, this
was already considered in ref. \cite{CV13} (where the focus was on results
with expanding space) for a duration of five periods. Over the final two
periods the rate of exponential decay appeared to diminish somewhat. There
$20\times20$ coarse-graining cells were used to evaluate $\bar{H}$. Here we
reconsider this case with a somewhat larger coarse-graining length,
corresponding to $16\times16$ cells (and with a new set of initial phases). We
provide more detailed plots and a best-fitting of the exponential decay. The
parameters used -- coarse-graining length, number of grid points per cell --
are the same as will be used in the case of four states below, to facilitate
comparison of the two cases.

A three-dimensional plot of our results is displayed in Figure 1. The smoothed
nonequilibrium density $\tilde{\rho}$ is shown in the left-hand column while
the smoothed equilibrium density $\tilde{\rho}_{\mathrm{QT}}$ is shown in the
right-hand column. The first row shows the densities at the initial time
$t=0$, while the third row shows the densities at the final time $t=10\pi$
(after five periods). The second row shows the densities at the intermediate
time $t=5\pi$.%

\begin{figure}
\begin{center}
\includegraphics[width=0.7\textwidth]%
{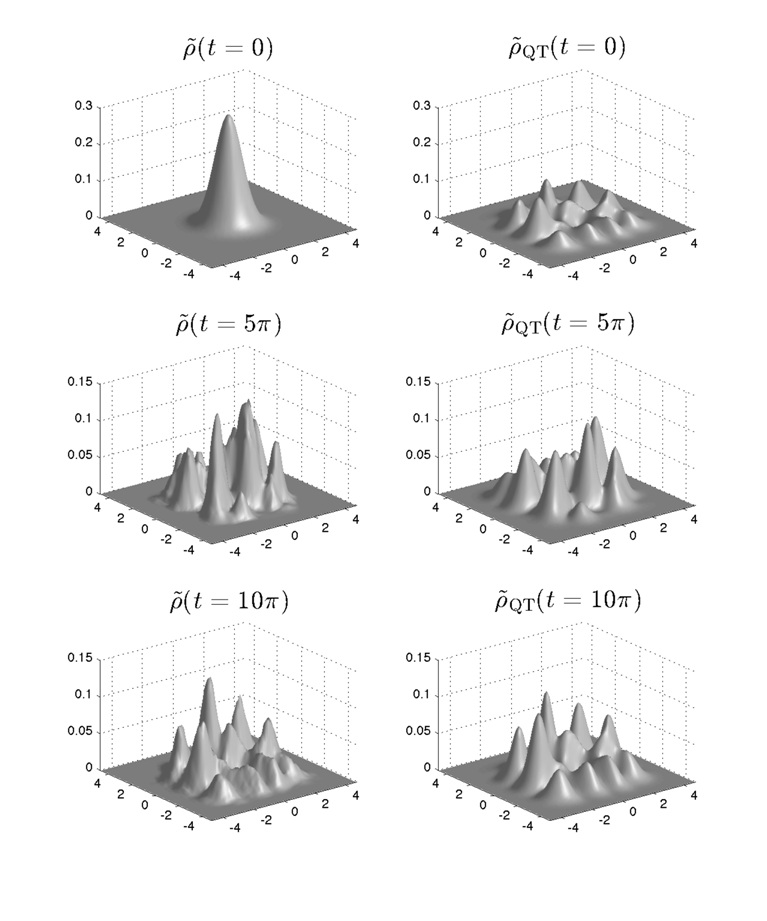}%
\caption{Relaxation for a case with 25 energy states. The wave function is
periodic with period $2\pi$. The left-hand column shows the (smoothed)
nonequilibrium density $\tilde{\rho}$. The right-hand column shows the
(smoothed) equilibrium density $\tilde{\rho}_{\mathrm{QT}}$. The first, second
and third rows show the densities at the respective times $t=0$, $5\pi$ and
$10\pi$ (up to five periods).}%
\end{center}
\end{figure}

As with previous simulations, by eye the relaxation towards equilibrium is
obvious. To quantify the relaxation we plot $\bar{H}(t)$ as a function of time
(in units of the period $2\pi$). This is shown in Figure 2, where values for
$\bar{H}$ are calculated and plotted at every period. The three separate runs,
with three different grids, yield slightly different results for $\bar{H}$
which are shown in different colours. In the displayed plot, the values
obtained for each run have been joined by straight lines. We also display (as
a dashed line) the best fit to an exponential function of the form%
\begin{equation}
a\exp[-b(t/2\pi)]+c\ , \label{exp}%
\end{equation}
with constant parameters $a$, $b$ and $c$. (We require that the best fit curve
passes through the value at $t=0$.)

In Figure 3 we plot $\ln\bar{H}$ against time. There we display error bars at
each time, whose limits are the extremal values of the three results for
$\ln\bar{H}$, and the centres of the error bars are joined by straight lines.
From Figure 3 we may see that the decay is approximately exponential. However,
the rate of decay during the final two periods is arguably slightly larger
than during the preceding three periods -- in contrast with the results of
ref. \cite{CV13} in which the decay rate was slightly smaller in the final two
periods. To within slight variations in the decay rate, we may say that for
$M=25$ we have confirmed an approximately exponential decay over five periods.
We also see that $\bar{H}$ drops from an initial value $\simeq1.26$ to a final
(mean) value $\simeq0.01$. The final value is less than $1\%$ of the initial
value. (The final value of $\bar{H}$ is significantly smaller than in ref.
\cite{CV13} because here the coarse-graining length is larger.\footnote{In the
limit of vanishing coarse-graining length $\bar{H}$ becomes equal to the
fine-grained value $H$ which remains constant in time \cite{AV91a,AV92,AV01}.})

As shown in Figure 2 the best-fit curve for $\bar{H}(t)$ has a small `residue'%
\begin{equation}
\bar{H}_{\mathrm{res}}=c\simeq0.02
\end{equation}
equal to about $2\%$ of the initial value. However, this is comparable to the
error (the differences in the three values obtained at four and five periods)
and so we are unable to say if the residue is real or significant.%

\begin{figure}
\begin{center}
\includegraphics[width=\textwidth]%
{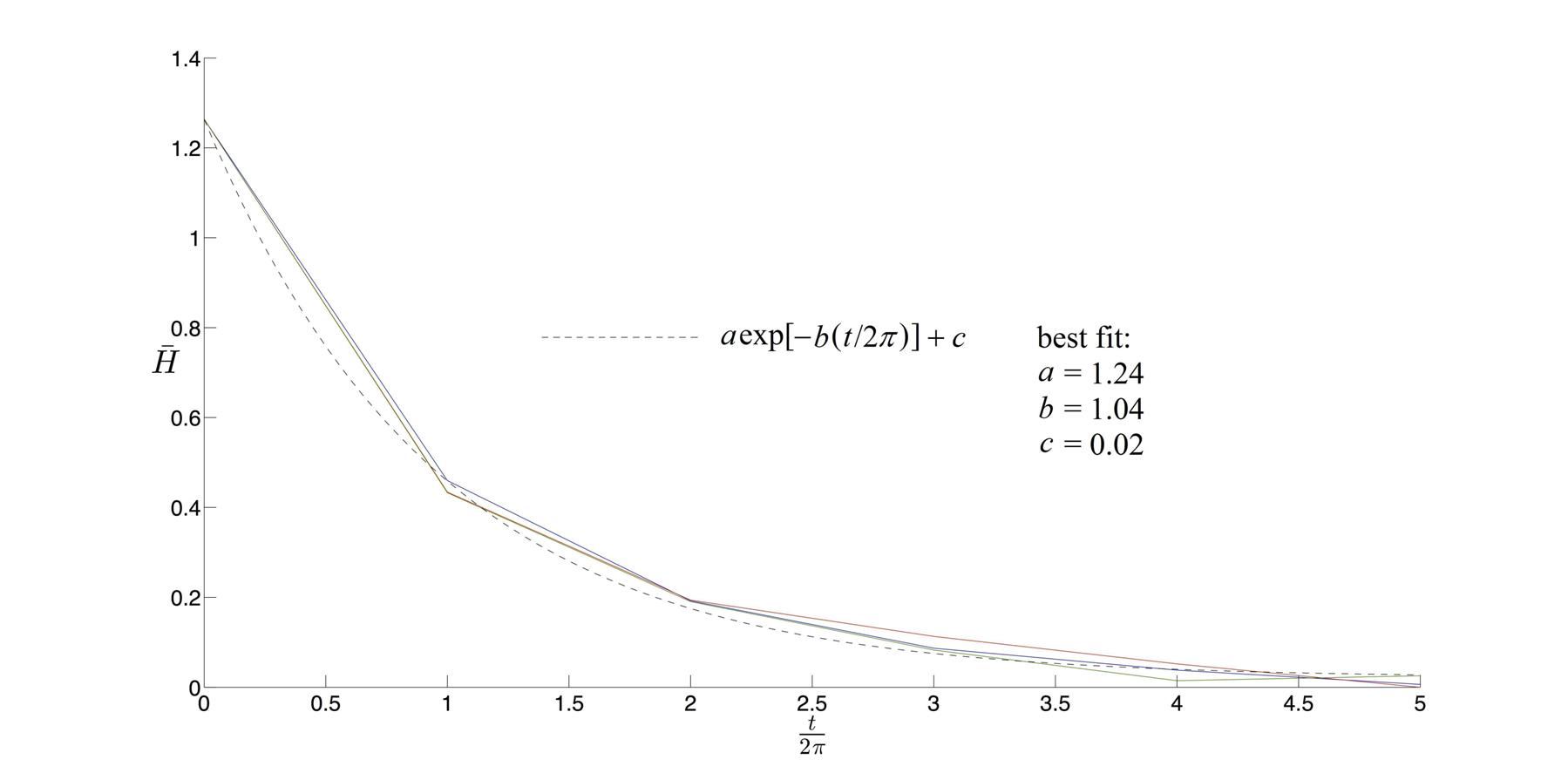}%
\caption{Plots of the coarse-grained $H$-function $\bar{H}(t)$ as a function
of time (in units of the period $2\pi$), for 25 energy states. Three separate
simulations with three different grids yield slightly different results for
$\bar{H}$ (shown in different colours). In the display, the values obtained
have been joined by straight lines and we include (dashed line) a best fit to
an exponential with a residue. The residue $c$ is comparable to the error.}%
\end{center}
\end{figure}
%

\begin{figure}
\begin{center}
\includegraphics[width=\textwidth]%
{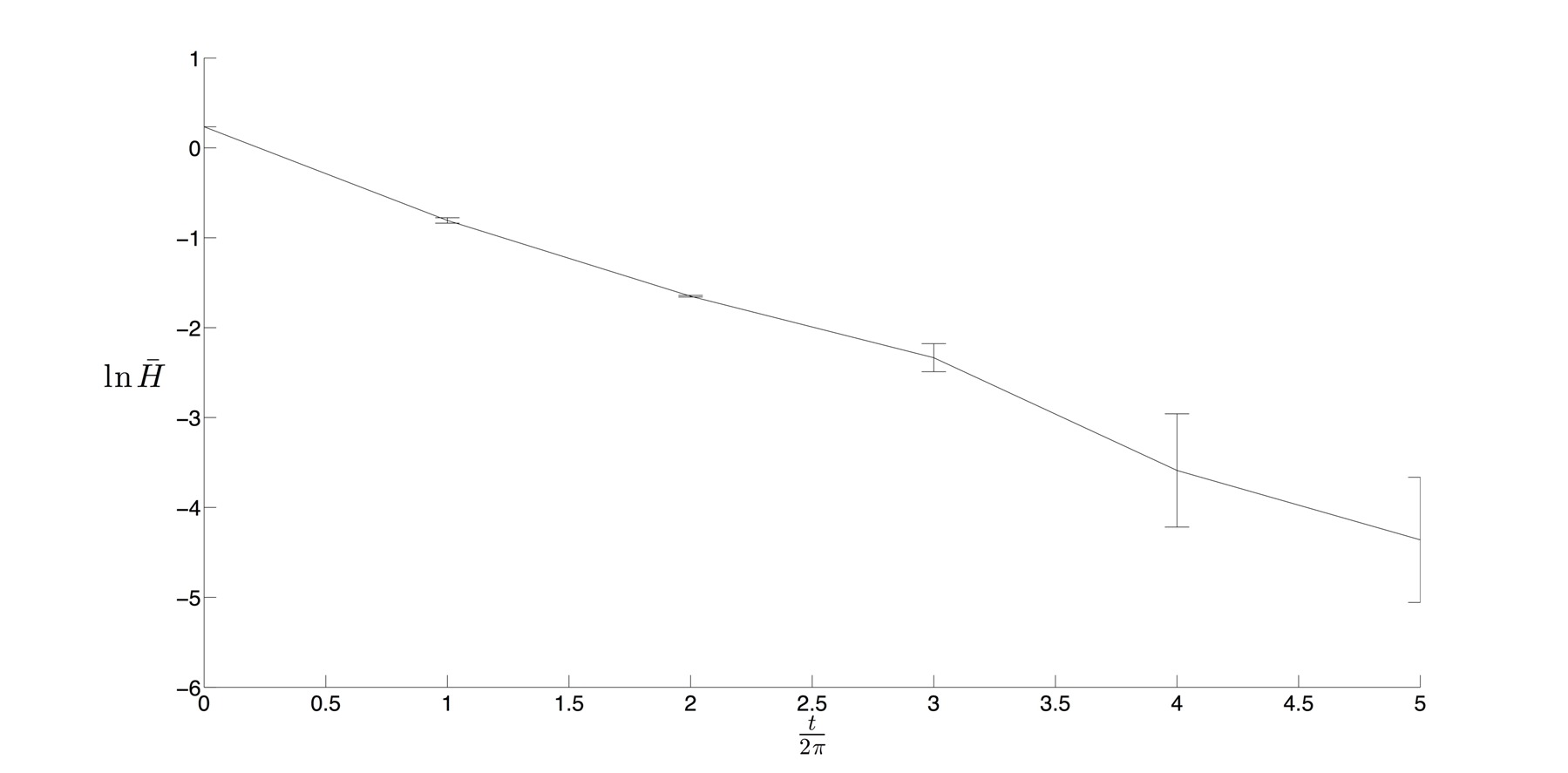}%
\caption{Plot of $\ln\bar{H}$ against time for a case with 25 energy states.
The error bars indicate the extremal values of the three results for $\ln
\bar{H}$. In the display the centres of the error bars are joined by straight
lines. The decay is approximately exponential.}%
\end{center}
\end{figure}

Does the decay continue indefinitely (whether exponentially or not), with
$\bar{H}$ asymptotically approaching zero? Or does the decay of $\bar{H}$ halt
at some point? We are unable to say, as we have been unable to calculate
$\bar{H}$ accurately beyond five periods. In Figure 2 the green curve
increases during the fifth period, suggesting that the decay of $\bar{H}$ may
be halting. But only further simulations over longer time intervals can settle
the matter. For $M=25$ the trajectories are very erratic and an accurate
computation of $\bar{H}$ beyond five periods has unfortunately proved to be
too difficult given our current methods. The difficulty lies not in the
computational time taken to perform the simulations, but in the excessive
number of inaccurate (discarded) trajectories that appear when we attempt to
carry the simulations further ahead in time. We have tried increasing the
maximal number of time steps per trajectory from $10^{7}$ to $10^{8}$ but with
little improvement in the percentage of accurate trajectories.

As things stand, we cannot confirm that the decay halts for this case with
$M=25$. However, we are able to confirm such a phenomenon for a case with
$M=4$.

\section{An example of long-time saturation of relaxation}

We now consider a case with just four energy states, $M=4$. The velocity field
is much milder and the trajectories less erratic. The numerical simulation is
therefore considerably easier and we are able to find accurate results for
$\bar{H}$ up to 50 periods (as opposed to just five periods for $M=25$%
).\footnote{For the record, to four decimals the initial phases $\theta_{mn}$
(as defined in (\ref{supn})) were taken to be: $0.5442$, $2.3099$, $5.6703$
and $4.5333$.}

A three-dimensional plot of our results is now displayed in Figure 4, where
again the smoothed nonequilibrium density $\tilde{\rho}$ is shown in the
left-hand column and the smoothed equilibrium density $\tilde{\rho
}_{\mathrm{QT}}$ is shown in the right-hand column. The first row shows the
densities at the initial time $t=0$, while the third row shows the densities
at the final time $t=100\pi$ (after 50 periods). The second row shows the
densities at the intermediate time $t=50\pi$.%

\begin{figure}
\begin{center}
\includegraphics[width=0.7\textwidth]%
{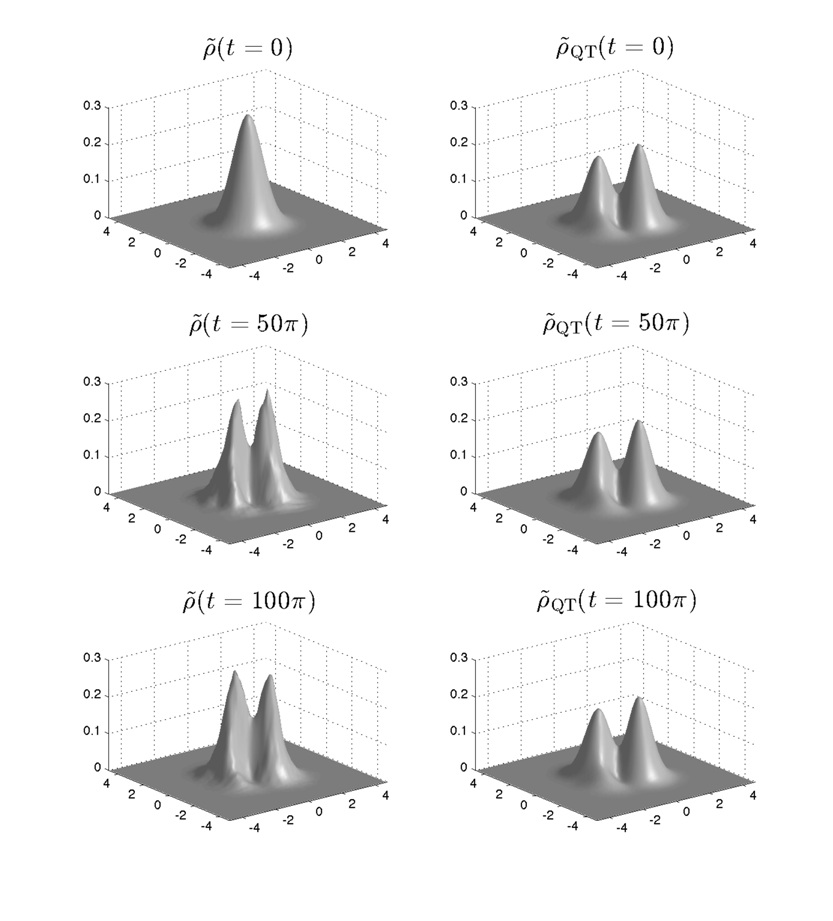}%
\caption{Saturation of relaxation for a case with four energy states. Again,
the left-hand column shows the (smoothed) nonequilibrium density $\tilde{\rho
}$ and the right-hand column shows the (smoothed) equilibrium density
$\tilde{\rho}_{\mathrm{QT}}$. The first, second and third rows show the
densities at the respective times $t=0$, $50\pi$ and $100\pi$ (up to 50
periods).}%
\end{center}
\end{figure}

Remarkably -- for such a tiny number of modes -- by eye one can still see a
clear relaxation towards equilibrium. The quantum state is of the very simple
form $\sim\left\vert 00\right\rangle +\left\vert 01\right\rangle +\left\vert
10\right\rangle +\left\vert 11\right\rangle $ (with randomly-chosen relative
initial phases). And yet, with such a minimal number of excitations above the
ground state, very significant relaxation still occurs if only one waits a
sufficiently long time. Furthermore, at least initially, the coarse-grained
function $\bar{H}(t)$ still decays approximately exponentially with time.

However, as shown in Figures 5 and 6 (where we respectively plot $\bar{H}$ and
$\ln\bar{H}$ against time), the decay of $\bar{H}$ is approximately
exponential only for the first 15 or 20 periods. (For the first 15 periods
values for $\bar{H}$ are calculated and plotted every period. Thereafter we
have one value at 20 periods and subsequently every 10 periods up until 50
periods. As before, at each time there are three different values of $\bar{H}$
obtained from three different grids, hence the error bars.)

Figure 5 again includes a best-fit to an exponential decay with a residue, of
the form (\ref{exp}). In this case the residue is significantly larger than
the error (the differences in the three values obtained, in particular at 40
and 50 periods). The residue appears to be real and significant.%

\begin{figure}
\begin{center}
\includegraphics[width=\textwidth]
{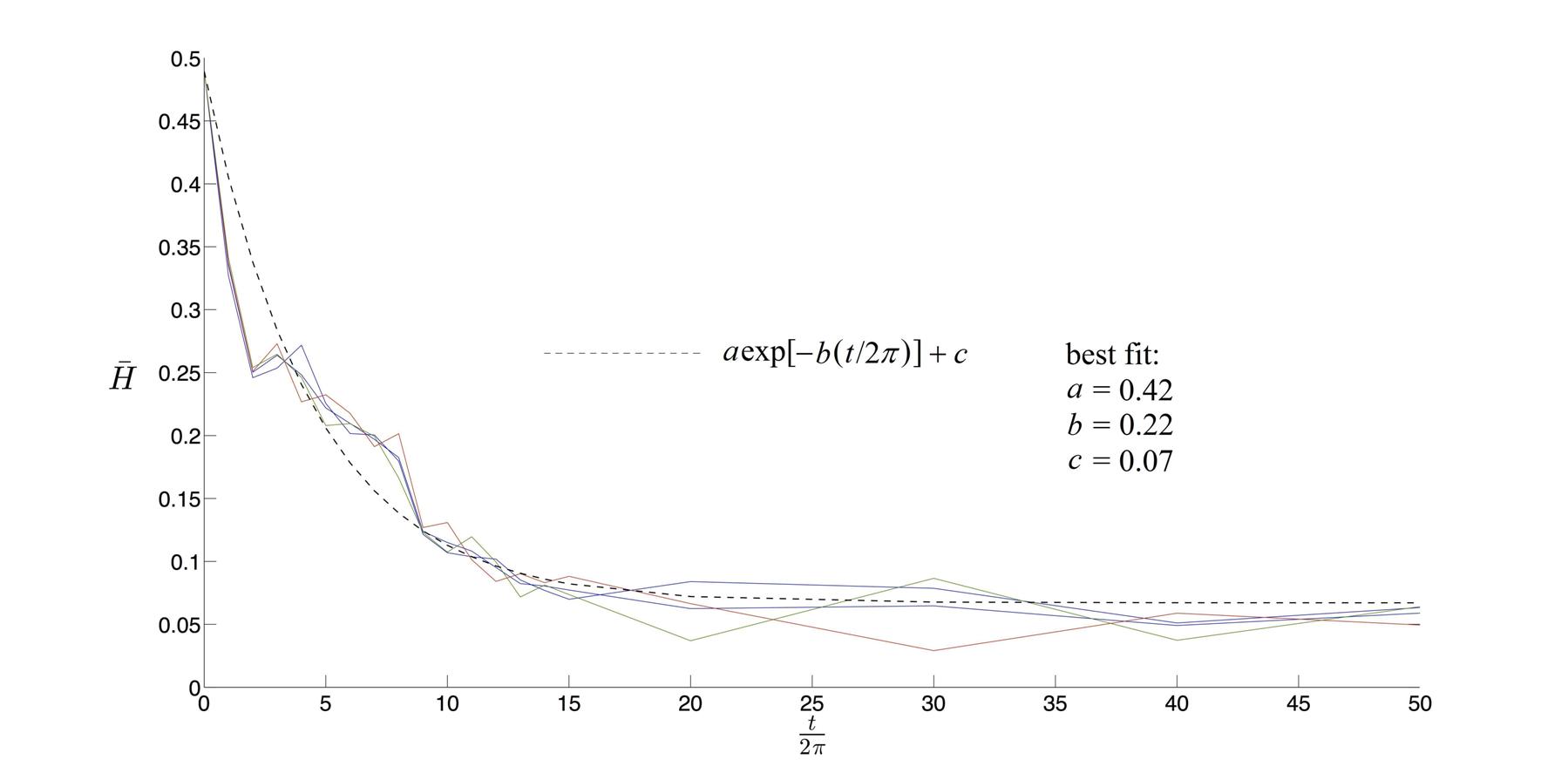}%
\caption{Plots of $\bar{H}(t)$ as a function of time for a case with just four
energy states. Again, three separate simulations yield slightly different
results for $\bar{H}$ and we include (dashed line) a best fit to an
exponential with a residue. The residue $c$ is now larger than the error. The
decay appears to saturate after about 20 periods ($t_{\mathrm{sat}}\simeq
40\pi$).}%
\end{center}
\end{figure}
%

\begin{figure}
\begin{center}
\includegraphics[width=\textwidth]
{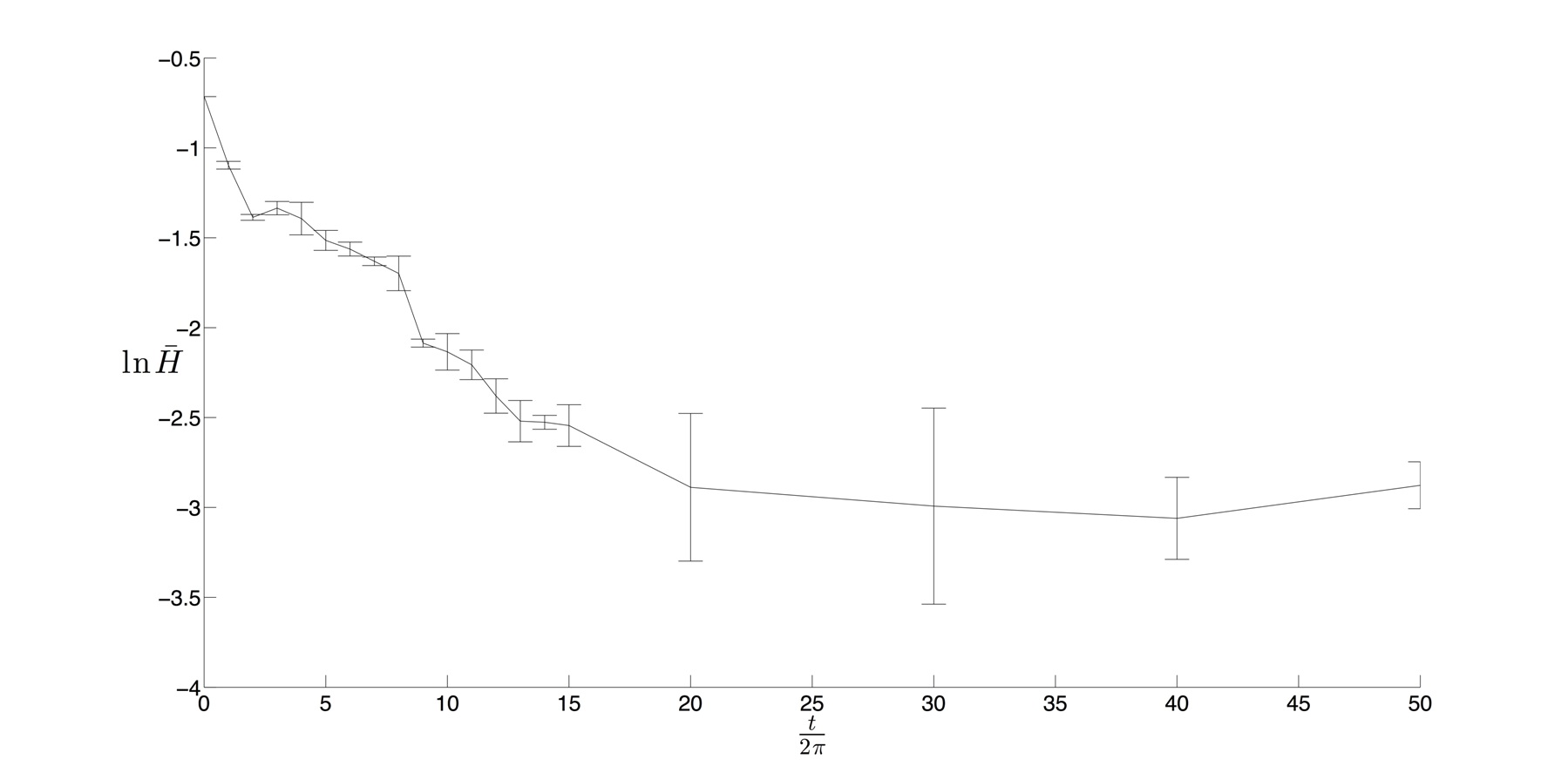}%
\caption{Plot of $\ln\bar{H}$ against time for a case with four energy states.
Again, the error bars indicate the extremal values of the three results for
$\ln\bar{H}$ and the centres of the error bars are joined by straight lines.
The decay is approximately exponential only for the first 15 or 20 periods.}%
\end{center}
\end{figure}

Thus, after about 20 periods ($t=40\pi$) the decay appears to halt: the
function $\bar{H}(t)$ levels off (roughly) to a constant residue%
\begin{equation}
\bar{H}_{\mathrm{res}}=c\simeq0.07\ .
\end{equation}
As we can see from Figure 5, $\bar{H}$ has dropped from an initial value
$\simeq0.49$ to a final (mean) value $\simeq0.06$ -- the final value is about
$12\%$ of the initial value. After as much as an additional 30 periods the
residue shows no sign of disappearing, and it seems reasonable to suppose that
no further relaxation (as measured by the decrease of $\bar{H}$) will take place.

For this case with four energy states, then, it appears that the exponential
decay levels off after about 15 or 20 periods, with a residue that exceeds
$10\%$ of the initial value. We may say that relaxation halts after about a
`saturation time'%
\begin{equation}
t_{\mathrm{sat}}\simeq40\pi\ .
\end{equation}

For all we know there may exist a (smaller) residue for the above case with
$M=25$ as well -- setting in at some time larger than the five periods we have
been able to calculate for. To settle this would require accurate simulations
over longer times for that case.

\section{Confinement of trajectories}

To understand these results, we study the `confinement' of the trajectories --
the tendency for them to remain within sub-regions of configuration space
instead of exploring the whole support of $\left\vert \psi\right\vert ^{2}$.
We use two simple but effective tests. First, we plot examples of single
trajectories obtained by evolving continuously forwards in time over many
periods. Second, we consider a selection of small (square)\ regions filled
with a large number of initial points, and we plot the final positions of
those points at a single final time (to show the `fate' of an initial square
after many periods). The first test gives a visual sense of the degree to
which trajectories explore the support of $\left\vert \psi\right\vert ^{2}$,
while the second test gives a visual sense of the degree to which neighbouring
initial points become widely scattered.

We first consider the case $M=25$, with the same initial wave function
(including phases) that was used in Section 3.

In Figure 7a we display trajectories obtained from a selection of 10 initial
points, evolved from $t=0$ to $t=10\pi$ (five periods).\footnote{The initial
points used are: (1.5,1.5), (1.5,--1.5), (--1.5,1.5), (--1.5,--1.5),
(0.5,0.0), (0.0,--0.5), (--0.5,0.0), (0.0,0.5), (0.25,0.25), (--0.25,0.25).}
In Figure 7b we display 10 initial squares at $t=0$ (each containing 100
points) and their fate at $t=10\pi$. (The squares are centred on the 10
initial points used previously.)%

\begin{figure}
\begin{center}
\includegraphics[width=0.8\textwidth]%
{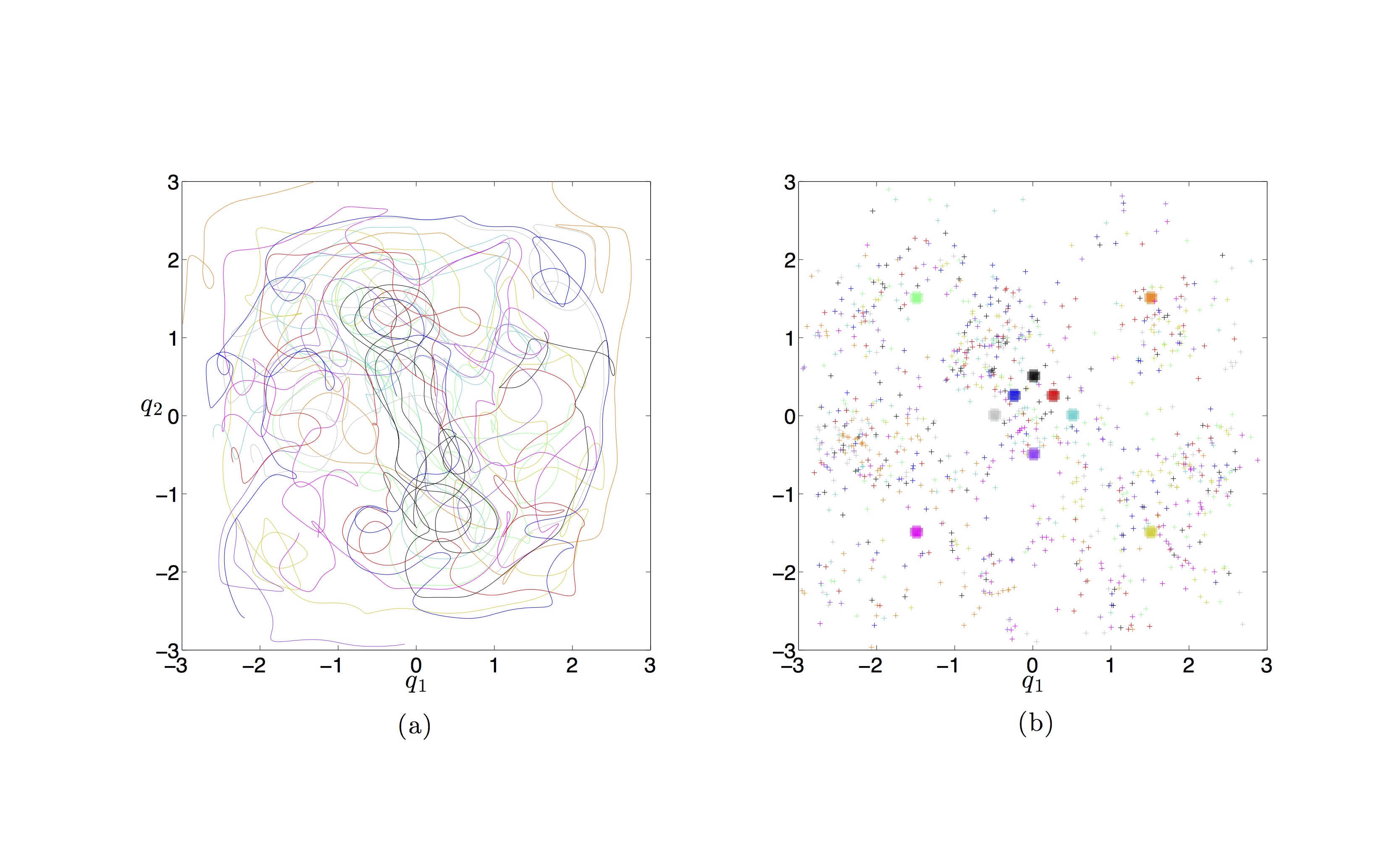}%
\caption{For the case $M=25$ (as in Section 3) we display (a) trajectories,
and (b) the fate of initial small squares, each calculated over five periods.
There is negligible confinement. (In (b), each of the initial squares yields
scattered final points only, with no `streaks' or other evident clustering.)}%
\end{center}
\end{figure}

The trajectories show negligible confinement. Instead, roughly speaking, they
tend to explore more or less the whole of the support of $\left\vert
\psi\right\vert ^{2}$. This is consistent with the absence of a discernible
residue in $\bar{H}$ for this case.

We now consider the case $M=4$, with the same initial wave function (including
phases) that was used in Section 4.

In Figure 8a we display trajectories obtained from the same selection of 10
initial points used above, now evolved from $t=0$ to $t=50\pi$ ($25$ periods).
In Figure 8b we display the same 10 initial squares at $t=0$ and their fate at
$t=50\pi$.%

\begin{figure}
\begin{center}
\includegraphics[width=0.8\textwidth]%
{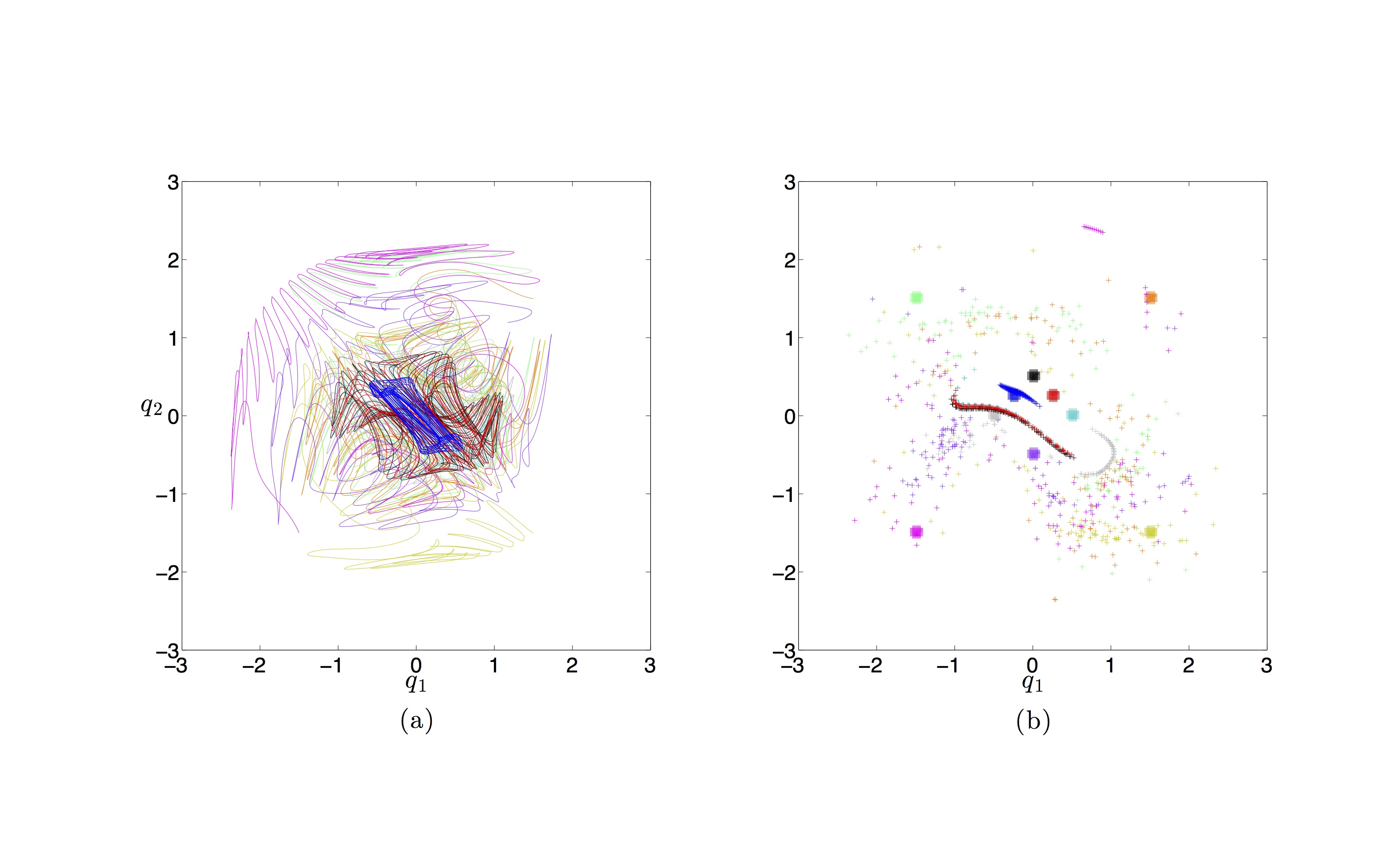}%
\caption{For the case $M=4$ (as in Section 4) we display (a) trajectories, and
(b) the fate of initial small squares, each calculated over $25$ periods.
There is strong confinement. (In (b), four squares yield clustered streaks
only (black, dark blue, red, light blue), one yields a streak and scattered
points (grey), five yield scattered points only (green, yellow, gold, purple,
magenta).)}%
\end{center}
\end{figure}

The trajectories now show a strong tendency to be confined. They do not
generally explore the whole of the support of $\left\vert \psi\right\vert
^{2}$. This is consistent with the presence of a large residue in $\bar{H}$
for this case.

So far we have only considered a particular choice of initial phases for each
case, $M=25$ and $M=4$. In the first case the trajectories tend to explore the
support of $\left\vert \psi\right\vert ^{2}$ and there is no discernible
residue in $\bar{H}$. In the second case the trajectories tend to be confined
and there is a definite large residue in $\bar{H}$. How common is this
behaviour for different choices of initial phases? Some straightforward tests
show that confinement is more likely to occur for low values of $M$ and less
likely to occur for high values of $M$ (assuming that the initial phases are
chosen randomly).

To see this we examine similar plots -- of trajectories and of the fate of
initial squares -- for different sets of randomly-chosen initial phases and
for varying values of $M$. Specifically, we have plotted trajectories obtained
from the same selection of ten initial points as above, but with ten different
sets of initial phases. This has been done for each of $M=4,\ 9,\ 16$ and $25$
(with final times $t=50\pi$, $20\pi$, $15\pi$ and $10\pi$ respectively). We
have also plotted the fates of the same ten initial squares as above, again
with ten different sets of initial phases and for each of $M=4,\ 9,\ 16$ and
$25$ (again with the listed final times).

For $M=4$ we find that about two thirds of cases -- where each case
corresponds to a different set of initial phases -- show strong confinement
(comparable to that seen in Figure 8) while about one third show only mild
confinement. An example of `mild' confinement is shown in Figure 9. For $M=9$
about half of the cases show strong to mild confinement; while for about half
of the cases confinement is negligible (comparable to that seen in Figure 7).
For $M=16$ about half of the cases show mild to low confinement; while for
about half of the cases confinement is negligible. For $M=25$, about one third
of the cases show mild confinement; for about two thirds of the cases
confinement is negligible.%

\begin{figure}
\begin{center}
\includegraphics[width=0.8\textwidth]%
{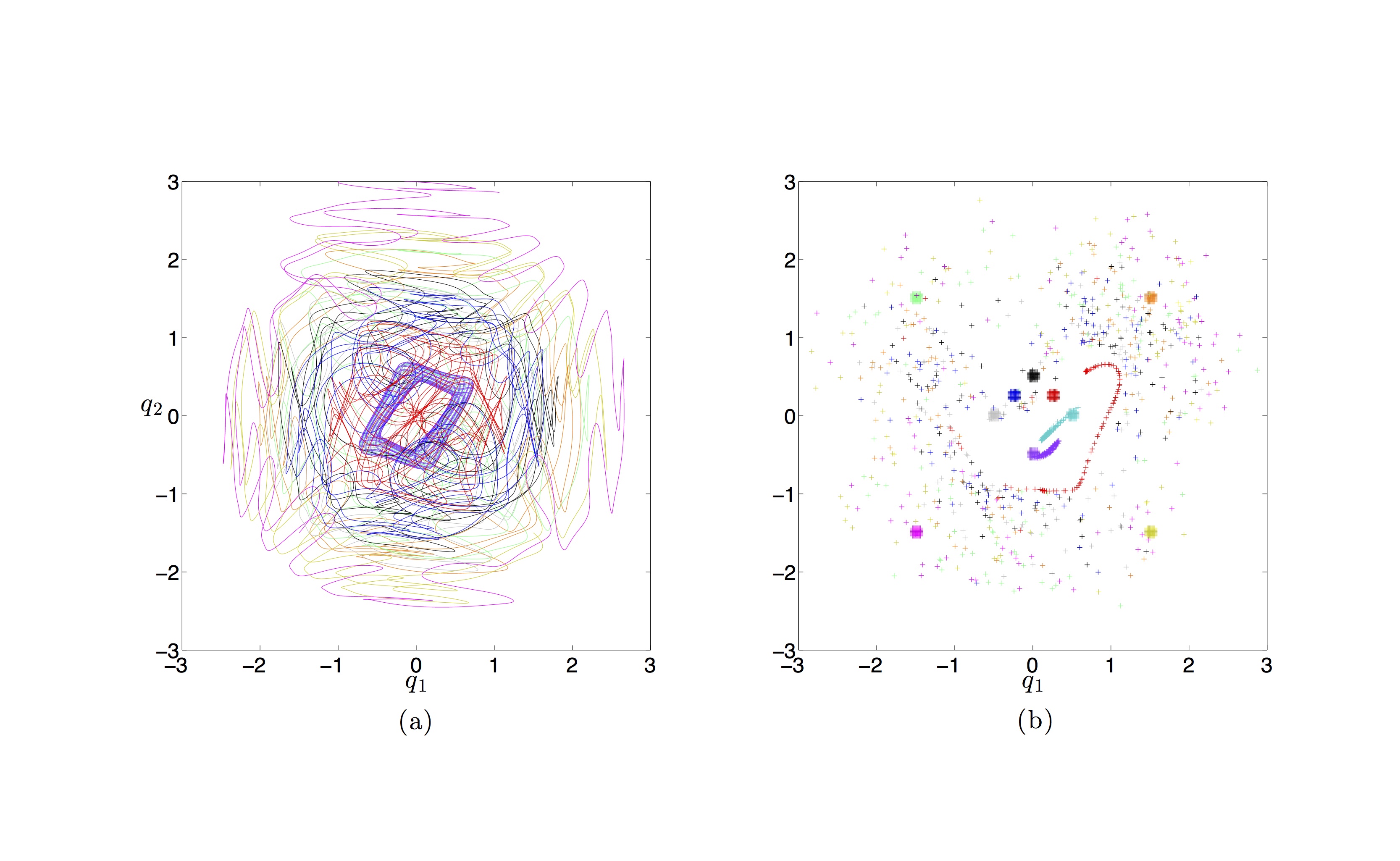}%
\caption{An example of mild confinement for $M=4$, showing (a) trajectories,
and (b) the fate of initial small squares, each calculated over $25$ periods.
(In (b), two squares yield clustered streaks only (light blue, purple), one
yields a long streak with some scatter (red), seven yield scattered points
only (green, magenta, grey, black, dark blue, yellow, gold).)}%
\end{center}
\end{figure}

We have not attempted to quantify the degree of confinement precisely. For the
present purpose it suffices to make a judgement by eye from the plots, which
show a clear tendency for stronger confinement at lower $M$ and milder or
negligible confinement at larger $M$.

For $M=4$, while most choices of initial phases yield strong confinement some
do not. In a latter case we would expect there to be a smaller residue in
$\bar{H}$ or perhaps no discernible residue at all. This has been confirmed by
running simulations for $\bar{H}$ as done in Section 4 but with a different
set of initial phases chosen such that the trajectories show only a small
amount of confinement.\footnote{The following set of phases was used (to four
decimals): $\theta_{11}=0$, $\theta_{21}=6.2782$, $\theta_{12}=2.0865$,
$\theta_{22}=0.2582$.} The results of our confinement tests for this case are
shown in Figure 10. We find no discernible residue in $\bar{H}$. After a time
evolution of $30$ periods, the value of $\bar{H}$ is indistinguishable from
zero (to within the accuracy of our simulations), as shown in Figure 11.%

\begin{figure}
\begin{center}
\includegraphics[width=0.8\textwidth]%
{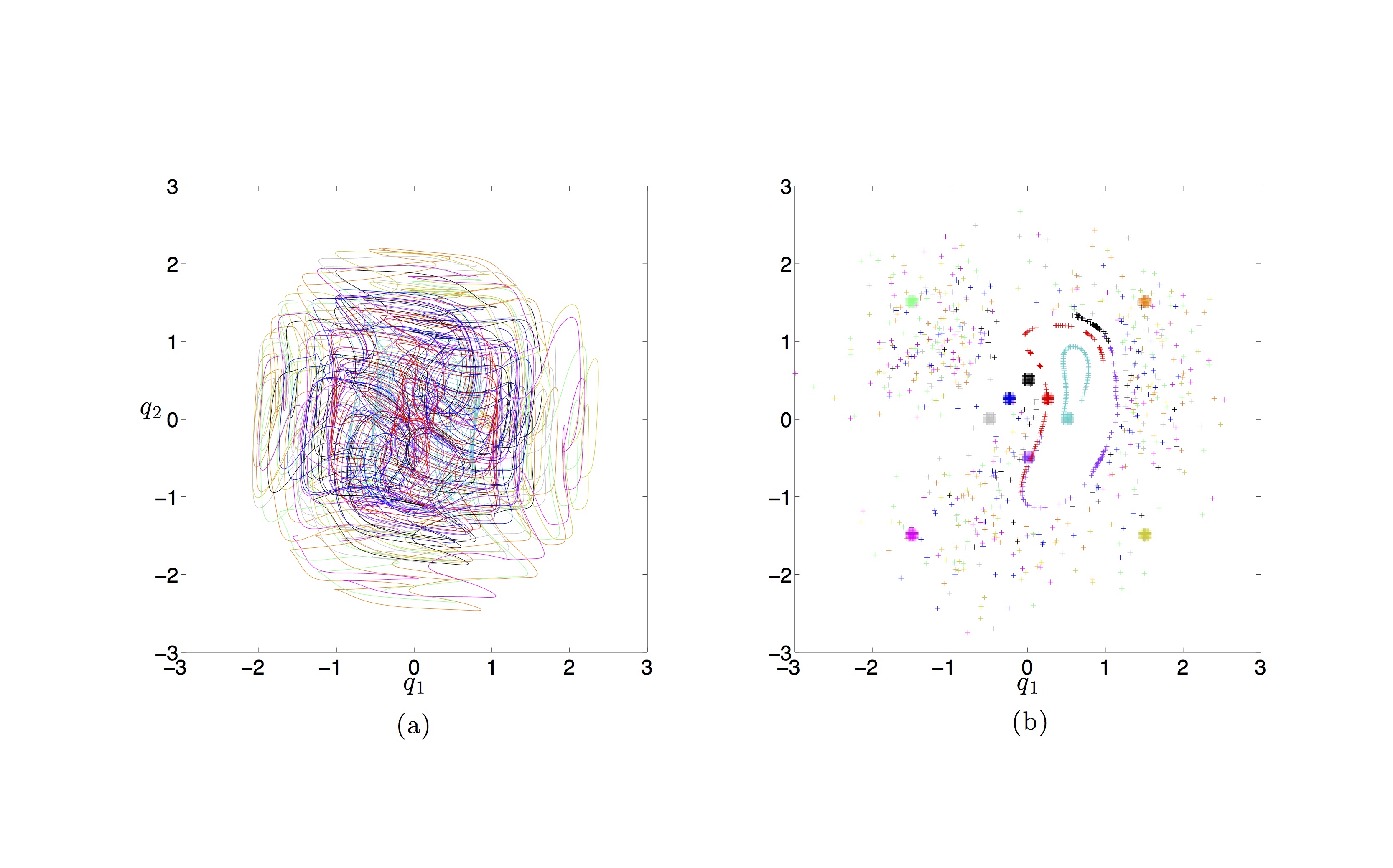}%
\caption{For $M=4$, with selected initial phases yielding only a small amount
of confinement (calculated over $25$ periods). (In (b), one square yields a
clustered streak only (red), three yield both streaks and scattered points
(black, light blue, purple), while six yield scattered points only (dark blue,
green, yellow, gold, magenta, grey).)}%
\end{center}
\end{figure}
%

\begin{figure}
\begin{center}
\includegraphics[width=0.8\textwidth]%
{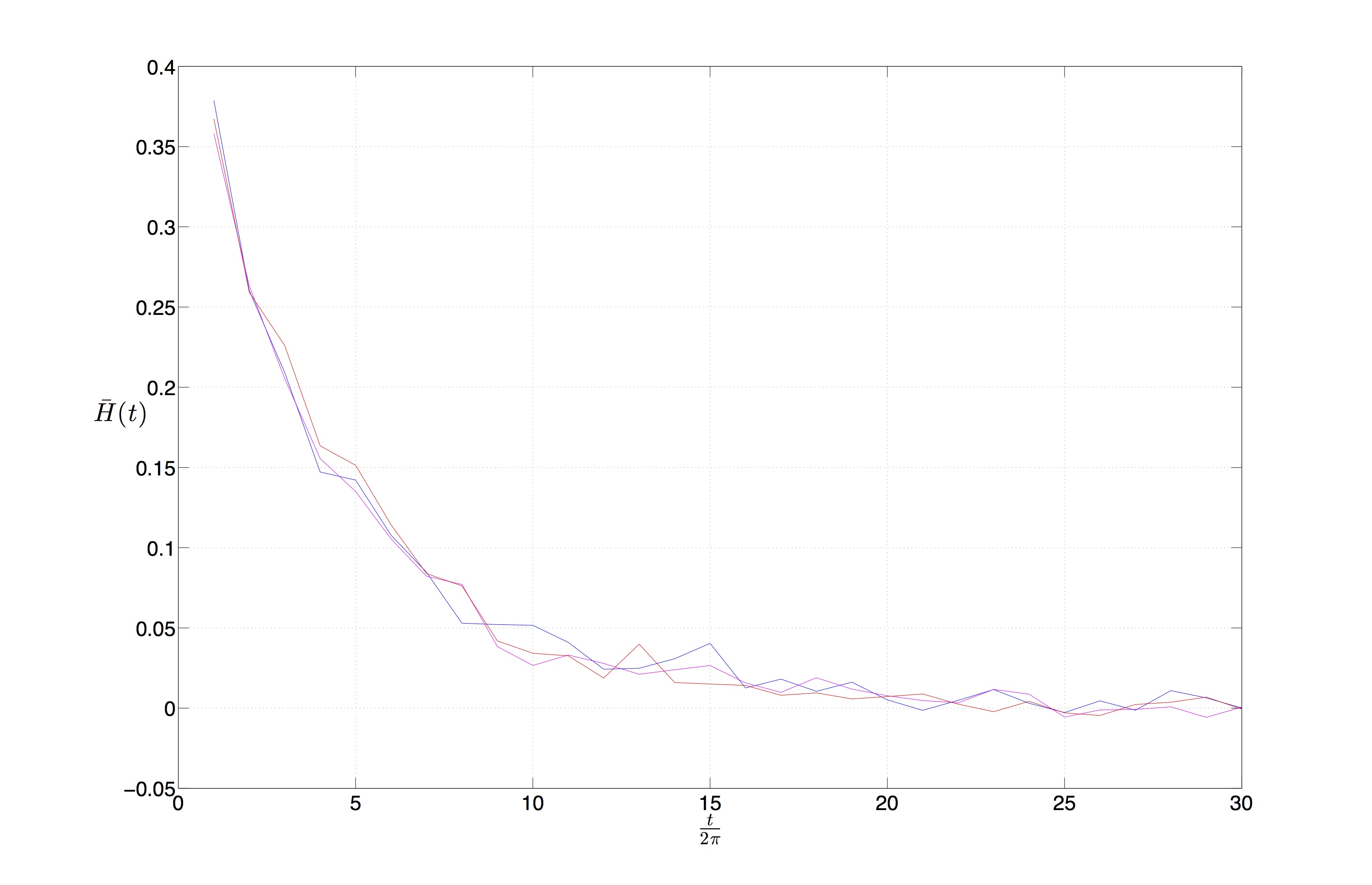}%
\caption{Simulations of $\bar{H}(t)$ for $M=4$, with a set of initial phases
chosen such that the trajectories show only a small amount of confinement.
After $30$ periods, there is no discernible residue in $\bar{H}$.}%
\end{center}
\end{figure}

Similarly, for $M=25$, while most choices of initial phases yield negligible
confinement some choices do not. In a case with significant confinement we may
expect to find a residue in $\bar{H}$.

While there is clearly room for further and more quantitative study, we may
draw the following broad conclusions. As the number $M$ of energy states
increases, we are less likely to find confinement of the trajectories; it is
more likely that the trajectories will explore the bulk of the support of
$\left\vert \psi\right\vert ^{2}$. (The measure of `likelihood' is defined by
choosing the initial phases randomly on the unit circle, where each choice of
phases corresponds to a choice of initial wave function.) A large or
significant residue in $\bar{H}$ is then more likely to exist for smaller $M$
and less likely to exist for larger $M$. A precise study of how the likelihood
diminishes with $M$ is left for future work.

\section{Conclusion}

For a superposition of 25 energy states we have confirmed an approximately
exponential decay of the coarse-grained $H$-function $\bar{H}(t)$ to a final
value that is less than $1\%$ of the initial value (after five periods of
periodic wave function evolution, the limit beyond which we were unable to
calculate $\bar{H}$ accurately). In contrast, for a case with just four energy
states we were able to calculate much further in time, and we found that the
approximately exponential decay of $\bar{H}(t)$ halts or saturates after a
time $t_{\mathrm{sat}}$ equal to about 20 periods or so, with a large residue
$\bar{H}_{\mathrm{res}}$ that exceeds $10\%$ of the initial value $\bar{H}%
(0)$. For the latter case we were able to calculate $\bar{H}$ accurately for
an additional 30 periods -- that is, up to a total of 50 periods -- and we
have confirmed that the residue shows no sign of disappearing. We are able to
explain these different results in terms of the extent to which the
trajectories explore the full support of $\left\vert \psi\right\vert ^{2}$. In
the first case the trajectories show negligible confinement to sub-regions of
configuration space, whereas in the second case the trajectories show strong
confinement. The lack of full exploration of the support of $\left\vert
\psi\right\vert ^{2}$ in the second case explains why we obtain a large
residue $\bar{H}_{\mathrm{res}}$. We also found a case with four energy states
for which the trajectories show only a small amount of confinement and
$\bar{H}(t)$ becomes indistinguishable from zero (to the accuracy of our
simulations). We have studied how common the two kinds of behaviour are likely
to be, for different wave functions with randomly-chosen initial phases. We
conclude that the confinement of trajectories -- and an associated large
residue $\bar{H}_{\mathrm{res}}$ -- are less likely to occur when the initial
wave function contains a larger number $M$ of energy states.

Many open questions remain. It may be that for any given $M$, no matter how
large, significant confinement and an associated large residue $\bar
{H}_{\mathrm{res}}$ always exist for some sets of initial phases -- though we
would expect such sets to become increasingly rare as $M$ increases. On the
other hand, it seems conceivable that if $M$ exceeds a critical value
$M_{\mathrm{crit}}$ there will be no significant confinement regardless of the
initial set of phases. We hope that further study will clarify this. Even in
the absence of significant confinement, there could still exist a
\textit{small} residue $\bar{H}_{\mathrm{res}}$ at sufficiently large times
$t_{\mathrm{sat}}$. For all we know this could be true for any finite $M$. If
so, we would like to know how $t_{\mathrm{sat}}$ and $\bar{H}_{\mathrm{res}}$
scale with $M$. Presumably $\bar{H}_{\mathrm{res}}$ will decrease with
increasing $M$, since the increased complexity of the velocity field makes the
trajectories more erratic and so will drive the system closer to equilibrium.
The same may be true for $t_{\mathrm{sat}}$. (Note that we have discussed
relaxation in terms of a coarse-graining approach for isolated systems
\cite{AV91a,AV92,AV01}, modelled on the analogous classical discussion
\cite{Tol, Dav}. If a residue $\bar{H}_{\mathrm{res}}$ does always exist for
any finite $M$, this might be viewed as an artifact of treating the system as
strictly isolated.) On the other hand, again, it seems conceivable that if $M$
exceeds some other critical value $M_{\mathrm{crit}}^{\prime}$ there will be a
continued exponential decay of $\bar{H}(t)$ asymptotically to zero. These are
tantalising open questions that require further study, perhaps with improved
numerical methods.

Finally, we briefly address possible cosmological consequences.

For ordinary systems emerging from a hot big bang, at early times there will
have been a huge number $M$ of energy states superposed in the relevant wave
functions, and so any residue $\bar{H}_{\mathrm{res}}$ today (if there is
such) is likely to be extremely small -- presumably corresponding to a
residual nonequilibrium at a very tiny lengthscale (as originally suggested in
ref. \cite{AV92}). Perhaps this lengthscale is so small, for the overwhelming
majority of sets of initial phases, as to be essentially unobservable. Only
improved simulations will be able to settle the matter. (In contrast, for
relic particles that decoupled sufficiently early, it is conceivable that
deviations from equilibrium exist on quite large scales today; see refs.
\cite{AV01,AV07,AV08}.)

On the other hand, our results for $M=4$ show that a large residue $\bar
{H}_{\mathrm{res}}$ is likely to exist if the number of excitations was always
sufficiently small. A system with such a small number of excitations would not
be expected to have ever existed at high temperatures in the early universe.
However, as we have mentioned, it is possible that there was a
pre-inflationary era with very few excitations above the vacuum. Our results
suggest that, even if such an era lasts a very long time, there is likely to
be a significant residual nonequilibrium during inflation at all relevant
wavelengths (and not only at long wavelengths owing to the suppression of
relaxation discussed in refs. \cite{AV07,AV08,AV10,CV13}).

\textbf{Acknowledgements}. A.V. is grateful to Patrick Peter for helpful
discussions. The work of A.V. and S.C. was funded jointly by the John
Templeton Foundation and Clemson University.

\end{document}